\documentstyle[12pt,aaspp4]{article}
\def\deg{\arcdeg}
\def\etal{{\it et al.}}
\def\eg{{\it e.g.,}}

\def\lya{Ly-$\alpha$}

\begin{document}

\title{High-Resolution $K'$ Imaging of the $z=1.786$ Radio Galaxy 3C\,294}

\author {Alan Stockton\altaffilmark{1}, Gabriela Canalizo\altaffilmark{1}}

\affil{Institute for Astronomy, University of Hawaii, 
2680 Woodlawn Drive, Honolulu, HI 96822}

\author {Susan E. Ridgway}

\affil{Department of Physics and Astronomy, Johns Hopkins University, 
Homewood Campus, Baltimore, MD 21218}

\altaffiltext{1}{Visiting Astronomer, Canada-France-Hawaii Telescope,
operated by the National Research Council of Canada, the Centre National de
la Recherche Scientifique de France, and the University of Hawaii.}

\begin{abstract}
We have obtained imaging in the $K'$ band ($\sim I$-band rest frame)
of the $z=1.786$ radio galaxy
3C\,294 with the 36-element curvature-sensing adaptive optics system 
{\it Hokupa`a} and the Canada-France-Hawaii Telescope.  At a resolution of
$\lesssim0\farcs15$, the galaxy is seen as a group of small but resolved
knots distributed over a roughly triangular region $\sim1\farcs4$ across.
The interpretation of the structure depends on the location of the nucleus,
as indicated by the compact radio core.  Its position is uncertain
by $\gtrsim0\farcs5$ ($2\sigma$) because of uncertainties in the optical
astrometry, but our best estimate places it at or near the southern apex
of the distribution.  If this location is correct, the most likely 
interpretation is that of a hidden quasar nucleus illuminating dusty
infalling dwarf-galaxy-like clumps having characteristic sizes of $\sim1.5$ kpc.
\end{abstract}


\section{Introduction}

High-redshift radio galaxies probably mark out regions of higher than
average density in the early Universe, and they give us a window into 
formation processes for at least one kind of massive galaxy.
Recent {\it Hubble Space Telescope} ({\it HST}) imaging of radio galaxies
with redshifts $>2$ in the rest-frame ultraviolet show that most comprise
several components and that the ultraviolet flux from these components
is apparently dominated by recent star formation (van Breugel \etal\ 
\markcite{vbr98}1998; Pentericci \etal\ \markcite{pen98}1998, 
\markcite{pen99}1999).

3C\,294, at $z=1.786$, 
is one of the dozen radio galaxies in the 3CR catalog with
$z>1.5$, so it is one of the most powerful radio sources in the
observed Universe as well as a likely example of a massive galaxy in
its youth.  It has strong \lya\ emission extending over $\sim10$\arcsec,
roughly aligned with the radio structure (McCarthy \etal\ 
\markcite{mcC90}1990).  It also has a $V=12$ star $<10$\arcsec\
to the west (Kristian, Sandage, \& Katem \markcite{kri74}1974), 
which initially hampered
optical and IR observations, but which now can be put to good
use as an adaptive optics (AO) reference.  In this {\it Letter} we describe
the results of an initial AO imaging investigation of 3C\,294.

\section{Observations and Data Reduction}

The AO observations were obtained with the University of Hawaii AO system
{\it Hokupa`a} (Graves \etal\ \markcite{gra98}1998) mounted at the f/36 focus 
of the Canada-France-Hawaii Telescope (CFHT).  This system uses the 
curvature-sensing approach pioneered by Fran\c{c}ois Roddier (Roddier,
Northcott, \& Graves
\markcite{rod91}1991).  Briefly, an image of the telescope primary is formed
on a 36-element deformable mirror.  Light from this mirror shortwards of 
$\sim1\ \mu$m is sent by a beamsplitter to a membrane mirror, which is
driven at 2.6 kHz to image extrafocal images on both sides of focus onto
a 36-element avalanche-photodiode array.  Corrections for the wavefront
errors derived from the difference of these extrafocal images are sent
back to the deformable bimorph mirror, which is updated at 1.3 kHz.
Under typical seeing conditions at CFHT, and for sufficiently bright 
($R\sim12$) stars, diffraction-limited imaging can be achieved as short
as $I$ band (Graves \etal\ \markcite{gra98}1998).  Our imaging of 3C\,294
was in the $K'$ band (Wainscoat \& Cowie \markcite{wai92}1992), so the
correction was excellent and quite stable over the course of the
observations.  The detector system was the University of Hawaii Quick 
Infrared Camera (QUIRC), which uses a $1024\times1024$ HAWAII array
(Hodapp \etal\ \markcite{hod96}1996).  
We obtained 8 300 s exposures on 1998 July 3 UT 
and 14 300 s exposures on 1998 July 4 UT.  Unusually rapid variation of
the airglow emission compromised the reduction of data from the first night,
and we use here only the data from the second night.  The individual
exposures were dithered in a pattern that kept the bright guide star off
the edge of the detector for all exposures.  The images were reduced using
our standard iterative procedure (Stockton, Canalizo, \& Close 
\markcite{sto98}1998).  Briefly, we make a bad-pixel mask from a
combination of hot pixels (from dark frames) and low-response pixels
(from flat-field frames); these pixels are excluded from all sums and
averages.  For each dark-subtracted frame, 3 or 4 time-adjacent
dithered frames were median averaged to make a sky frame, which was
normalized to the sky value of the frame in question and subtracted; then the
residual was divided by a flat-field frame.  These first-pass images
were registered to the nearest pixel and median averaged.  After a
slight smoothing, this rough combined frame was used to generate object
masks for each frame, which were then combined with the bad-pixel mask.
This process was repeated to give better sky frames, better offsets for
alignment, and a new combined image.  This new image was used to replace
bad pixels in each flattened frame with the median from the other frames,
so that bad pixels would not affect the centering algorithm used to
calculate the offsets.  The final combined image was a straight average
of the corrected, subpixel-registered frames, using a sigma-clipping
algorithm in addition to the bad-pixel mask to eliminate deviant pixel
values.  Normally, we use field stars for registration, but in this
case we had to use the ghost image of the guide star, since there were no other
objects visible on individual frames.  This ghost image was produced by a
secondary reflection in the AO system optics; it was about 10 mag
fainter than the guide star, and it was positioned 0\farcs114 south and 
7\farcs85 east of the star.  The image scale with the $K'$ filter was
determined to be $0\farcs03537 \pm 0\farcs00005$ pixel$^{-1}$, based on an
accurate measurement of the scale in the $J$ filter by Claude Roddier and
a determination of the ratio of the $K'$ to $J$ scales for the QUIRC camera
from contemporaneous imaging data obtained with the University of Hawaii 2.2 m 
telescope.

\section{Results}

Our $K'$ band image of 3C\,294 samples the
rest-frame spectrum from 6850--8300 \AA, a region unlikely to be dominated
by emission lines (the strongest expected lines in this bandpass are
[\ion{Ar}{3}] $\lambda\lambda7136$,7751 and [\ion{O}{2}] 
$\lambda\lambda7320$,7330).  This region is also close to the peak of
the spectral-energy distribution (SED) for a stellar population with an
age of $\gtrsim2$ Gyr.  If the luminosity at the center of this galaxy is
dominated by a central bulge of relatively old stars, we should be
able to see it.  What we actually do see is shown in Fig.~1.
Scattered light and a diffraction spike
from the guide star extend from the lower right across the lower middle of
the frame, and the ghost image of the guide star lies just above the
diffraction spike.  3C\,294 is at the middle of the frame.  The structure is
knotty and filamentary, comprising several distinct components 
within a roughly triangular region about 1\farcs4 ($\sim10$ kpc, for
$H_0=75$, $q_0=0.3$, $\Lambda=0$, which we assume throughout this {\it Letter})
across, no one of which is clearly dominant.  The data from the first night,
although of lower quality, confirm the main features seen in Fig.~1.

Since, at larger scales, both the inner radio structure and the extended
Ly-$\alpha$ emission are aligned along PA$\sim20$\deg\ (McCarthy \etal\
\markcite{mcC90}1990), one possible interpretation is that we are seeing 
a dust scattering nebula centered on this same axis 
(Chambers \markcite{cha99a}1999$a$,\markcite{cha99b}$b$).  If this is the 
case, the hidden nucleus must lie somewhere near the faint southern tip of the
``triangle''.  However, it is also possible that we are seeing an
assemblage of small merging components, and that the active nucleus
is coincident with one of the peaks in our image.  Thus, the interpretation
of the observed structure is critically dependent on the location of the 
nucleus.
 
McCarthy \etal\ \markcite{mcC90}(1990) found a flat-spectrum central compact
radio component, presumably to be identified with the nucleus, and they
determined its position to a precision of $\sim50$ mas.  Because we have
determined the position of ghost image and the image scale quite accurately,
we can relate any point in our field
to the position of the guide star to a similar precision.  However,
in order to relate our image to the radio position, we need to have
high-quality astrometric positions for both the guide star and the radio
nucleus in the same reference frame.  There are two problems with 
accomplishing this task:  (1) the McCarthy \etal\ \markcite{mcC90}(1990) 
radio data is 
referenced to the equinox B1950, epoch 1979.9 VLA calibrator reference
frame, and converting to the equinox J2000, epoch 2000.0 FK5 frame is
not completely straightforward; and (2) as McCarthy \etal\ 
\markcite{mcC90}(1990)
point out, the positions for our guide star given by Veron 
\markcite{ver66}(1966), Kristian \etal\ \markcite{kri74}(1974), and 
Riley \etal\ \markcite{ril80}(1980) do not agree very well.  Our 
separate AO imaging of the guide star itself shows an added minor 
complication:  the star is a double with
a separation of 0\farcs13 (see inset in Fig.\ 1).  

We have converted the position of the central radio component to the FK5
frame by an empirical mapping of the B1950 VLA calibrator reference frame
in the region of the source to the J2000 VLA calibrator frame.  We first
do a standard precession of B1950 calibrator positions from equinox B1950
to nominal equinox J2000, for calibrators within 15\arcdeg\ of 3C\,294.  
We then take the difference between these precessed
positions and the cataloged J2000 VLA calibrator positions, and we fit a 
low-order
reference surface to these differences (which were always 0\farcs5 or
less).  The values of the corrections
at the position of 3C\,294 are then applied to the precessed B1950 VLA
position for the central compact radio source.  We obtain
a position $14^{\rm h} 06^{\rm m} 44\fs077\pm0\fs002, +34\arcdeg11\arcmin
24\farcs956\pm0\farcs025$ (J2000), where the quoted errors are the 
standard errors 
of the fit of the correction surface to the coordinate residuals.
In addition, there is an uncertainty due to errors in the J2000 VLA
calibrator positions, which is likely to be $\sim0\farcs1$ or less.

Unfortunately, the AO guide star is too faint to appear in the 
Hipparcos/Tycho databases.  In order to obtain a more accurate position
for it, we have obtained large-image-scale, short-exposure CCD frames 
covering an 8\arcmin\ field including the star and
have derived a plate solution from 12 USNO-A2.0 stars (after deleting 3 
additional stars
with outlying residuals).  The residuals for the solution give a
standard error of 0\farcs24 in RA and 0\farcs34 in declination, so the
2$\sigma$ uncertainty in the position of an individual star is over
0\farcs5.  For the mean position of the components of the AO guide star, we
obtain $14^{\rm h} 06^{\rm m} 43\fs35, +34\arcdeg11\arcmin
24\farcs2$ (J2000).

The USNO-A2.0 catalog is in the FK5 system tied to the Hipparcos 
reference frame.  With the AO guide
star and 3C\,294 radio core positions, and our knowledge of the position of
the AO guide star relative to its ghost image, we can locate the position
of the radio core on our image.  This position, with $2\sigma$ error bars
(determined mostly by the uncertainty of the position of the guide star),
is shown in the lower-left inset in Fig.~1, along with radio-core
positions based on positions for the guide star given previously by 
Veron \markcite{ver66}(1966), Kristian \etal\ \markcite{kri74}(1974), 
and Riley \etal\ \markcite{ril80}(1980).
The position we derive for the radio core is consistent with a location
at or near the southern tip of the $K'$ structure, and apparently
inconsistent with association with any of the three brightest
blobs along the northern edge of the structure.

In a 3\arcsec-diameter synthetic aperture, we obtain a total $K'$ magnitude
for 3C\,294
of $18.3\pm0.3$.  Lower-resolution imaging at the NASA Infrared Telescope
Facility gives $K'=18.3\pm0.1$.  Both of these measurements are consistent
with the $K=18.0\pm0.3$ found by McCarthy \etal\ \markcite{mcC90}(1990) for a similar aperture.

\section{Discussion}

Radio galaxies at high redshifts often show multiple components, and these are
often found to be aligned with the radio structure (\eg\ Pentericci \etal\ 
\markcite{pen99}1999, and references
therein).  This conclusion is based mostly on optical data, which corresponds
to rest-frame UV for these objects, although recently some deep ground-based
IR imaging has been presented by van Breugel \etal\ \markcite{vBr98}(1998), 
and programs
involving HST NICMOS imaging of high-redshift radio galaxies are in progress
(\eg\ Fosbury \etal\ \markcite{fos99}1999).
One reason that 3C\,294 is especially of interest in this
context is that its redshift places the $K'$ band in a region of the 
spectrum that is unlikely to be
dominated by emission-line radiation or nebular thermal continuum. Since
there is also no evidence in the VLA map of 
McCarthy \etal\ \markcite{mcC90}(1990) of radio 
counterparts to the rest-frame optical structure we see, the emission is 
most likely due either to stars or to scattering from a hidden quasar.
The position we obtain for the flat-spectrum radio component of 3C\,294 
favors a location at or near the southern apex of the
observed distribution of bright knots, and it therefore supports (though
it cannot prove) the interpretation of the observed structure as an 
illumination cone, most likely due to dust scattering of radiation from
a quasar nucleus.  Figure 1 can be compared with Fig.~2 of McCarthy
\etal\ \markcite{mcC90}(1990), which shows contours of the Ly-$\alpha$ and
radio emission (the short arrows in our Fig.~1 are at the positions of the
radio knots K$_{\rm N}$ and K$_{\rm S}$ shown in their figure).  
The extended Ly-$\alpha$ emission also shows a well-defined triangular 
structure extending to the north, which McCarthy \etal\ \markcite{mcC90}(1990) 
suggest is due to anisotropic emission of ionizing
radiation from a central non-thermal source.  While the inferred Ly-$\alpha$
and $K'$ cones are fairly well aligned, the Ly-$\alpha$ material extends
$\sim4$\arcsec, or roughly 4 times as far as does the continuum emission
in our $K'$ AO image.  The Ly-$\alpha$ emission also extends to the other
side of the radio nucleus, in a weak and somewhat poorly-defined ``counter
cone.''  We do not see a corresponding feature at $K'$, but our dynamic
range is not sufficient to put very strong limits on the presence of
such material.  If the illumination is truly biconical and intrinsically
fairly symmetric, the southern cone must suffer significant extinction along
our line of sight.

However, granting that the morphology we see in our AO image is likely
to be at least partly determined by illumination effects, the material 
being illuminated also does
seem to have an intrinsic distribution in small ($\sim0\farcs2\approx1.5$ kpc)
coherent bodies, which (on the not unreasonable assumption that they 
contain stars as well as dust) must shortly merge together.
In fact, the tendency of these objects to show elongation roughly aligned in the
direction of the apex may well be due to tidal stretching.
Thus, the alternatives of illumination effects and merging subunits need not 
be starkly opposed to each
other in this case, although the presence of an illumination
cone, if confirmed (say, by polarization measurements), means that we are likely
seeing only part of the action.

In summary, 3C\,294 appears to be a particularly good example of several 
aspects of an emerging picture of high-redshift radio galaxies.  
The observations suggest the following scenario:
A quasar nucleus, hidden along our line of
sight, is responsible for the jets that power the radio source as well
as for the illumination of material in the immediate environment of the
radio galaxy that falls within a biconical region.  This illumination
is made evident to us by scattering by dust and by emission from 
the large Ly-$\alpha$ nebula that is aligned with the radio
axis (McCarthy \etal\ \markcite{mcC90}1990).  
The bulge is apparently still in the process of being assembled from small
($\sim1.5$ kpc = 0\farcs2), merging, dusty objects (\eg\ Baron \& White
\markcite{bar87}1987; Pascaralle \etal\ \markcite{pas96}1996), which, 
at the depth of our current images,
are visible only by scattered light, when they happen to fall within the
illumination cone of the central source.  Deeper high-resolution imaging
might pick up intrinsic emission from similar objects in regions not
illuminated by the quasar.

\acknowledgments

We are grateful for the support of the University of Hawaii Adaptive Optics
group:  Fran\c{c}ois Roddier, Claude Roddier, Buzz Graves, Malcolm Northcott,
and Laird Close, without which these observations would not have been possible.
We also thank Laird Close and Claude Roddier for helpful discussions and 
for information on the image scale, Dave Monet and Dave Tholen for
discussions on astrometric matters, and Rob Whitely for obtaining short
exposures on the 3C\,294 field.  This research was supported in part by 
NSF grant AST 95-29078.

\newpage

\begin{figure}[htp]
\caption{Image of 3C\,294 (centered) in the $K'$ filter obtained with the 
University of Hawaii AO system Hokupa`a at the CFHT.  The wings and
diffraction spike from the AO guide star (placed outside the field of the
detector) are visible to the right.  The compact bright spot just below
and to the left of 3C\,294 is a $10^{-4}$-intensity ghost image of the guide
star produced by the optics of the AO system.  The FWHM of the image is
$\sim0\farcs15$.  The upper right inset shows 3C\,294 at lower contrast,
and the upper-left inset shows a short exposure on the AO guide star, 
which is found to be
double, with a separation of 0\farcs13 and an intensity ratio at $K'$ of
1.5:1.  The large lower-left inset reproduces
the region around and to the east of 3C\,294, with estimates of the
position of the radio nucleus shown.  The large cross indicates the position 
and 2$\sigma$ internal error range of the radio core based on a position for
the AO guide star from an astrometric fit to 12 USNO-A2.0 stars.
The three ``$\times$'' points show the position of the radio core from positions
for the AO guide star given by Kristian \etal\ (1974), Veron (1966), and
Riley \etal\ (1980) (left to right, respectively).  Assuming that the radio
core actually lies near the southern apex of the optical structure, the
short arrows in the main panel are at the locations of the inner knots 
of the radio jets and 
point towards the directions of the hot spots, as seen on the 6 cm VLA map
of McCarthy \etal\ (1990).
}
\end{figure}

\end{document}